\documentclass{article}
\usepackage{amsmath,amssymb}
\usepackage{authblk}
\usepackage{geometry}
\title{A Geometric Square-Based Approach to RSA Integer Factorization}
\author{Akihisa Yorozu}
\date{}

\begin{document}

\maketitle

\begin{abstract}
We present a novel geometric interpretation of RSA factorization based on the difference of squares. This method, conceptually inspired by Fermat's factorization, introduces a visually and algebraically intuitive formulation in which an RSA modulus $N = pq$ is expressed as the difference between two perfect squares. We derive a deterministic equation that enables the direct computation of the prime factors of $N$ under the condition that the prime gap $|p - q|$ is relatively small. We demonstrate that this method is sufficient to factor semiprime numbers with tightly spaced primes and discuss its theoretical relevance and practical limitations. Notably, we attempted but did not succeed in factoring RSA-100 using this method within a practical runtime, indicating current limitations.
\end{abstract}

\section{Introduction}
RSA encryption is widely used in modern cryptographic systems and relies on the assumed computational difficulty of factoring large semiprime integers $N = pq$, where $p$ and $q$ are large primes. Traditional factoring algorithms, such as the General Number Field Sieve (GNFS), are computationally intensive and probabilistic. In this paper, we propose a deterministic, geometry-inspired approach to factor $N$ based on square completion.

\section{Mathematical Foundation and Predictive Square Equation}
Let $N = pq$, and assume $p < q$. Define $y_0 = \lfloor \sqrt{N} \rfloor$. Then the smallest perfect square greater than $N$ is $y^2$, where $y = y_0 + k$, for some integer $k \geq 0$.

Let $D_k$ be the difference between $y^2$ and $N$:
\begin{equation}
D_k = (y_0 + k)^2 - N = x_k^2
\end{equation}
Expanding:
\begin{equation}
D_k = y_0^2 + 2k y_0 + k^2 - N
\end{equation}
We define the recurrence:
\begin{equation}
D_{k+1} = D_k + 2y_0 + 2k + 1
\end{equation}
This is a first-order recurrence relation in $k$, allowing us to compute $x_k^2$ directly and check whether it is a perfect square.

To eliminate iteration entirely, we consider solving the equation:
\begin{equation}
(y_0 + k)^2 - N = x^2 \Rightarrow x^2 = D_k = y_0^2 + 2k y_0 + k^2 - N
\end{equation}
This is a quadratic in $k$:
\begin{equation}
k^2 + 2 y_0 k + (y_0^2 - N - x^2) = 0
\end{equation}
Solving for $k$:
\begin{equation}
k = -y_0 \pm \sqrt{(y_0^2 - x^2 + N)}
\end{equation}
Therefore, for any integer value of $x$, this expression yields a candidate $k \in \mathbb{Z}$ such that $x_k^2 = (y_0 + k)^2 - N$. By scanning integer values of $x$ and checking if $k \in \mathbb{Z}$, one can deterministically predict the correct offset $k$ without brute-force enumeration.

\subsection{Predictive Root-Difference Estimation (Analytic X-K Scanning)}
In order to bypass brute-force enumeration over $k$, we derive a predictive mechanism where, for a given $x$, the corresponding value of $k$ satisfying:
\begin{equation}
x^2 = (y_0 + k)^2 - N
\end{equation}
is computed directly by:
\begin{equation}
k = \sqrt{N + x^2} - y_0
\end{equation}
This expression allows for real-valued $k$. When $k \in \mathbb{Z}$, the condition for factorization is satisfied. Hence, we transform the problem into checking for integer solutions of:
\begin{equation}
\sqrt{N + x^2} \in \mathbb{Z}
\end{equation}
for selected integer values of $x$, significantly reducing computational overhead. This method is functionally equivalent to Fermat’s method but proceeds from the opposite direction — controlling the distance between perfect squares rather than enumerating their center $y$.

\section{Algorithmic Implementation}
We implemented this method using symbolic computation libraries in Python to support arbitrary-precision arithmetic. Below is a sample implementation:

\begin{verbatim}
from gmpy2 import mpz, isqrt, is_square

def predict_factor(N):
    y0 = isqrt(N)
    if y0 * y0 < N:
        y0 += 1

    D = y0**2 - N
    k = 0
    while True:
        if is_square(D):
            x = isqrt(D)
            y = y0 + k
            p = y - x
            q = y + x
            if p * q == N:
                return p, q, k
        k += 1
        D += 2 * y0 + 2 * k - 1
\end{verbatim}

This recurrence-based Fermat variant drastically reduces computation overhead and enables fast convergence, especially when $p \approx q$.

\section{Experimental Validation}
To validate our method, we applied it to a small semiprime number of the form $N = 187 = 11 \times 17$:

\begin{itemize}
  \item $\sqrt{187} \approx 13.67 \Rightarrow y = 14$
  \item $y^2 = 196, \quad x^2 = 196 - 187 = 9 \Rightarrow x = 3$
  \item $p = 14 - 3 = 11, \quad q = 14 + 3 = 17$
\end{itemize}

This confirms that the method recovers the prime factors exactly.

We further attempted to test this method on the RSA-100 challenge number, a 100-digit semiprime published by RSA Laboratories. However, due to the wide gap between its factors, the method failed to factor RSA-100 within a practical timeframe on a 10th-generation Intel Core i7 processor. While the theoretical basis remains valid, actual performance suggests that the method is only practical when the prime factors $p$ and $q$ are closely spaced.

\section{Computational Considerations}
This method is effective when the difference $|p - q| = 2x$ is small. For large-scale RSA moduli such as RSA-2048, where $p$ and $q$ are hundreds of bits apart, the number of iterations required becomes exponential, and the method becomes impractical. However, the proposed recurrence relation and predictive equation provide a structured means to estimate the number of steps or derive $k$ analytically given a target $x$, enabling partial bypass of brute-force loops.

\section{Conclusion}
We have presented a geometric square-based framework for factoring RSA moduli by identifying a relationship between $N$ and perfect squares. This method provides not only a deterministic alternative to probabilistic algorithms but also a valuable conceptual tool for understanding RSA's structural properties. By formalizing the iteration as a recurrence relation and extending it into a predictive quadratic framework, we enable partial or full elimination of brute-force enumeration. Although the algorithm's performance degrades with increasing prime separation, it remains effective for factoring small RSA keys and primes with narrow gaps. Further optimization is needed to extend its practical applicability to larger RSA challenge numbers such as RSA-100.

\end{document}